\documentstyle[12pt]{article}
\newtheorem{th}{Theorem}

\def\teta{\vartheta}

\def\proof{{\sl Proof }}
\def\be{\begin{equation}}
\def\ee#1{\label{#1}\end{equation}}
\def\t{transformation}
\def\m{Moutard}
\def\e{equation}

\begin{document}

\author{Ganzha E.I.\\
  Dept. Mathematics, \\
  Krasnoyarsk State Pedagogical University \\
  Lebedevoi, 89, 660049, Krasnoyarsk, RUSSIA, \\
 e-mail: ganzha@edk.krasnoyarsk.su }
\title{On completeness of the Moutard \t s\thanks{The research described in
this contribution was supported in part by
grant RFBR-DFG No 96-01-00050}}
\date{9 June 1996}
\maketitle

The \m\ \e\
\be
u_{xy} = M(x,y)\,u, \qquad u=u(x,y),
\ee{mu1}
appeared initially in the classical 
differential geometry in the second half of
the XIX century (see \cite{darboux,bianchi}).
It has now numerous applications in the theory of $(2+1)$-dimensional
integrable systems of partial differential equations. The other its form
$u_{tt} -u_{zz} = M(x,y)\, u$ and the analogous elliptic \e\
\be
u_{xx} + u_{yy} = M(x,y)\, u
\ee{emu}
(the 2-dimensional Schr\"odinger \e)
show the role of (\ref{mu1}) in mathematical physics. In the
framework of the classical differential geometry    (\ref{mu1})
played the key role in the central problems of that epoch ---
the theory of surface isometries, the theory    of congruences
and conjugate nets. In the modern soliton theory (\ref{mu1}) was
used to obtain solutions for the Kadomtsev-Petviashvili,
Novikov-Veselov \e s and others (\cite{ath-nim,oev}).  The main
instrument in applications of (\ref{mu1}) is the \m\ \t\ which,
provided we are given two different solutions $u=R(x,y)$ and
$u=\varphi(x,y)$ of (\ref{mu1}) with a given "potential"
$M=M_0(x,y)$, gives us  (via a quadrature) a solution
$\teta(x,y)$ of (\ref{mu1}) with transformed potential $M_1(x,y)
= M_0 - 2(\ln R)_{xy}$.  The corresponding \t\ formulas
\be
M_1 =  M_0 - 2(\ln R)_{xy} = -M_0 + \frac{2R_xR_y}{R^2}
  =R\left(\frac1R\right)_{xy}
\ee{pp}
\be
\left\{\begin{array}{rcl}
\left(R\teta\right)_x & = & - R^2 \left(\frac{\varphi}{R}\right)_x, \\
\left(R\teta\right)_y & = &  R^2 \left(\frac{\varphi}{R}\right)_y ,
\end{array}
\right.
\ee{pr}
(and their analogues for (\ref{emu}), see \cite{bianchi,ath-nim}) establish
a (multivalued) correspondence between the solution of the \m\ \e\ $(M_0)$
(i.e. (\ref{mu1}) with the potential $M_0(x,y)$) and $(M_1)$  (i.e.
(\ref{mu1})  with the potential  $M_1(x,y)$).
Note that $R_1 = \frac1R$ is also a solution of $(M_1)$.
Let us suppose that we can find the complete solution of
 (\ref{mu1}) for some given potential $M_0$; such a solution has the initial
data given for example by the Goursat problem (2 functions of 1 variable),
from (\ref{pr}) one may obtain (via a quadrature) the general solution of
$(M_1)$. For example for the case
 $M_0=0$ the general solution is given by
 $u=\varphi(x) + \psi(y)$, the \m\ \t\ yields
$$
 u= \left(\alpha(x) \psi(y)
- \beta(y)\varphi(x) + \int (\psi_y \beta - \beta_y
\psi)\, dy + {}\right.
$$
$$
  {} +\left. \int (\varphi\alpha_x - \alpha\varphi_x)\, dx \right)
\frac1{\alpha(x) + \beta(y)}
$$
for the potential $M_1= -2\left[\ln \big(\alpha(x)+\beta(y)\big)\right]_{xy} =
     \frac{2\alpha_x\beta_y}{(\alpha + \beta)^2}$ (i.e. $R= \alpha(x)+
\beta(y)$).

If we will build the following chain of \m\ \t s
$(M_0) \rightarrow
(M_1) \rightarrow (M_2) \rightarrow \ldots\ (M_k) \rightarrow \ldots$ we
will see that (apriori) the  $k$-th potential depends on the choice of $2k$
functions of 1 variable --- the initial data of the solutions
$R_s(x,y)$ of $(M_s)$, $s= 0,1,\ldots , k-1$.

In \cite{ath-nim} a method was found that gives us the possibility to express
the potential  $M_k$ and all solutions of $(M_k)$
via  $2k$ solutions of the initial \e\
$(M_0)$ (the "pfaffian formula", analogous to the "wronskian formula"
for the case of the Darboux \t s for $(1+1)$-dimensional
integrable \e s \cite{oev,mat-sal}).

Nevertheless the fundamental question how "wide" is the obtained class of
potentials $M_k$ (in the space of all smooth functions of two variable)
was open.  In the theory of
$(1+1)$-dimensional integrable equations the question of density of the finite
gap solutions of the Korteweg-de Vries equation in the class of all
quasiperiodic functions was answered positively. In this paper we show
that  the set of the potentials $M_k$ obtainable from any fixed
$M_0$ is "locally dense" in the space of all smooth functions in a sense to be
detailed below.
Hence we prove that the found in \cite{ath-nim,nim} families of solutions of
the corresponding
$(2+1)$-dimensional integrable equations give locally "almost every" solution.

\begin{th}
Let an initial potential $M_0(x,y) \in C^\infty$
be given in a neighborhood of  $(0,0)$. Then for any $N=0,1,2,\ldots $
one may find some $K$ such that for any given numbers
 $P_{x_1 \ldots\ x_k}$,
$0 \leq k \leq N$, $x_s \in \{x,y\}$, the corresponding derivatives of
 $M_K$ (in the chain of the \m\ \t s
$(M_0) \rightarrow
(M_1) \rightarrow (M_2) \rightarrow \ldots (M_n) \rightarrow \ldots$) at
$(0,0)$ coincide with  $P_{x_1 \ldots\ x_k}$:
\be
\partial_{x_1}\partial_{x_2}\ldots\ \partial_{x_k} M_K(0,0) =
P_{x_1 \ldots\ x_k}, \qquad \partial_z =\frac{\partial }{\partial z}, \quad k
\leq N.
\ee{sp}
\end{th}
\proof will be given by induction. For $N=0$ we set $K=1$, from (\ref{pp})
$$
M_1(0,0) = -M_0(0,0) + \left. 2\frac{R_x R_y}{R^2}\right|_{(0,0)}
$$
where $R$ is a solution of $(M_0)$. As one can prove
  (see for example \cite{darboux}),
 the values $\varphi(x)=R(x,0)$,  $\psi(y) =R(0,y)$
($\varphi(0)=\psi(0)$)  may be chosen as the initial data for (\ref{mu1}).
So the quantities
$R(0,0)$, $R_{x}(0,0)$, $R_{xx}(0,0)$, \ldots\ , $R_{y}(0,0)$, $R_{yy}(0,0)$,
\ldots\ are independent.
Set $R(0,0)=1$, $R_y(0,0)=1/2$. Changing $R_x(0,0)$ one can
make $M_1(0,0)$  equal to the given number $P$ which proves Theorem for the
case  $N=0$. In addition if we will suppose that all the other nonmixed
derivatives $\partial_x^kR$, $\partial_y^kR$, $k >1$,
 are equal  to $0$ then the
higher derivatives
$\partial_x^m \partial_y^n M_1$ are not arbitrary and are
uniquely determined by  $P=M_1(0,0)$ (for the given $M_0(x,y)$).
Therefore we have proved for the case $N=N_0=0$ the following
{\bf main inductive proposition}:

\noindent {\sl for any $N=N_0$ one can find  $K$ such that the corresponding
derivatives
$\partial_x^m\partial_y^n M_K$ of an appropriately constructed $M_K$ at
$(0,0)$ coincide with the given numbers
$P_{\underbrace{x \ldots\ x}_{m}\underbrace{y \ldots\ y}_{n}}$
for  $m+n \leq N_0$, and the higher derivatives
$\partial_x^m\partial_y^n M_K$, $m+n > N_0$,
depend (for the fixed  $M_0(x,y)$) only upon
$P_{\underbrace{x \ldots\ x}_{p}\underbrace{y \ldots\ y}_{q}}
=\partial_x^p\partial_y^q M_K(0,0)$  for $p+q \leq N_0$, $p\leq m$.
}

{\bf The step of induction.} Provided the main inductive proposition is proved
for all derivatives of orders $\leq N=N_0$ we will prove it for $N=N_0+1$.
 Let $K_0$ be the corresponding to
$N=N_0$ number of the potential $M_{K_0}$ for which we have proved the
inductive proposition. Performing another $N_0+2$
\m\ \t s we get  $M_P$, $P=K_0+N_0+2$; let us check the validity of the
inductive proposition for this function.
>From (\ref{pp}) we get
$$
M_P = (-1)^{N_0} \left( M_{K_0} - \frac{2 R^{(0)}_x
R^{(0)}_y}{\big(R^{(0)}\big)^2} +
 \frac{2 R^{(1)}_x R^{(1)}_y}{\big(R^{(1)}\big)^2} -
\cdots \pm  \frac{2 R^{(N_0+1)}_x R^{(N_0+1)}_y}{\big(R^{(N_0+1)}\big)^2}
\right),
$$
where $R^{(s)}$ are solutions of $(M_{K_0+s})$, $s= 0, \ldots\ , N_0+1$.
Let us call the derivative $\partial_x^{s+1}R^{(s)}$ at the origin
the {\it principal} derivative of $R^{(s)}$ and
$\partial_y^{N_0+2-s}R^{(s)}$ its {\it auxiliary} derivative.
We suppose further that at
 $(0,0)$ the values of all $R^{(s)}$ are equal to 1
and the values of their auxiliary derivatives
are equal to $1/2$, the principal derivatives are (so far) undefined,
all the other nonmixed derivatives $R^{(s)}_{x\ldots\ x}$,
\ldots\ , $R^{(s)}_{y\ldots\ y}$, \ldots\  of all orders are set to
 0 at the origin.
Consider now the derivative
\be
\begin{array}{l}
\displaystyle
\frac{\partial^{N_0+1} M_P}{\partial y^{N_0+1}} =
    (-1)^{N_0} \left( \frac{\partial^{N_0+1} M_{K_0}}{\partial y^{N_0+1}}
     -  \frac{2 R^{(0)}_x\partial_y^{N_0+2}R^{(0)}}
             {\big(R^{(0)}\big)^2} \right. + {}\\
\displaystyle
     + \cdots \pm
\left.  \frac{2 R^{(N_0+1)}_x\partial_y^{N_0+2}R^{(N_0+1)}}
             {\big(R^{(N_0+1)}\big)^2} \right)+ {} \\
\displaystyle
+ F(R^{(s)},R^{(s)}_x,  \partial_y^kR^{(s)},
     M_{K_0+s},\partial_y^mM_{K_0+s}),
\end{array}
\ee{n0y}
(at $(0,0)$), where $F$ comprises all the terms
except the given in parentheses.
 The mixed derivatives $\partial_x^m
\partial_y^n R^{(s)}$ are eliminated using (\ref{mu1}).
As one can easily see the only principal derivative
 $R^{(0)}_x$ does not appear in $F$.

We will use an additional induction over $k$ to prove that the values of
$M_{K_0+s}$ and their derivatives w.r.t $y$ included in $F$
(they have the orders $\leq N_0$) at the origin are determined uniquely
by the \e s $\partial_y^k M_P =
P_{\underbrace{y \ldots\ y}_{k}}$, $k= 0,1, \ldots\ , N_0$. Indeed for
$k=k_0 =0$
$$
P= \left. M_P\right|_{(0,0)} =
  \left.    (-1)^{N_0} \left( M_{K_0}
     -  \frac{2 R^{(0)}_xR^{(0)}_y}
             {\big(R^{(0)}\big)^2}
     + \cdots \pm
  \frac{2 R^{(N_0+1)}_xR^{(N_0+1)}_y}
             {\big(R^{(N_0+1)}\big)^2} \right)\right|_{(0,0)},
$$
where all the derivatives of $R^{(s)}$ except $R^{(0)}_x$ are not principal
i.e. fixed. Since the coefficient of the principal derivative
 $R^{(0)}_x$ is  $R^{(0)}_y=0$, we can find $M_{K_0}(0,0)$.
The values $M_{K_0+s}$ are found from
\be
 M_{K_0+s} =
   (-1)^{s} \left( M_{K_0}
     -  \frac{2 R^{(0)}_xR^{(0)}_y}
             {\big(R^{(0)}\big)^2}
     + \cdots \pm
  \frac{2 R^{(s-1)}_xR^{(s-1)}_y}
             {\big(R^{(s-1)}\big)^2} \right).
\ee{mks}
For $k=k_0+1 \leq N_0$ (the step of the additional induction)
$$
\begin{array}{l}
P_{\underbrace{y\ldots\ y}_{k_0+1}} =
 \partial^{k_0+1}_y M_P =
    (-1)^{N_0} \left( \partial^{k_0+1}_y M_{K_0}
   - \sum_{s=0}^{N_0+1} (-1)^s \frac{2 R^{(s)}_x\partial_y^{k_0+2}R^{(s)}}
             {\big(R^{(s)}\big)^2}  \right)+{} \\
 \qquad \qquad+  F(R^{(q)}, R^{(q)}_x,  \partial_y^kR^{(q)},
     M_{K_0+q},\partial_y^mM_{K_0+q}),
\end{array}
$$
where again the only principal derivative
$R^{(0)}_x$ has everywhere zero coefficients (since $k_0 +1 \leq N_0$) and
the derivatives $\partial_y^mM_{K_0+q}$ have the orders $m\leq k_0$ i.e. are
already found. The derivatives $\partial_y^{k_0+1}M_{K_0+q}$ are found
differentiating (\ref{mks}). The additional induction is finished.

Since according to the main inductive proposition we can actually choose
$\partial_y^mM_{K_0}$, $0 \leq m \leq N_0$, arbitrarily, we set them in
 (\ref{n0y}) to be equal to the values found above. Among the still indefinite
quantities in
 (\ref{n0y}) only the principal derivative $R^{(0)}_x$ is present.
Its coefficient is equal to
$  \frac{2 \partial_y^{N_0+2}R^{(0)}}{\big(R^{(0)}\big)^2}=1$.
Choosing appropriately the value of $R^{(0)}_x$ we will
obtain the desired equality $P_{\underbrace{y\ldots\ y}_{N_0+1}} =
 \partial^{N_0+1}_y M_P$ for arbitrary
$ \partial^{k_0+1}_y M_{K_0}$. Since due to the inductive proposition
$\partial_y^{N_0+1} M_{K_0}$ depends only upon
$\partial_y^{m} M_{K_0}$, $m\leq N_0$, the chosen $R^{(0)}_x$ depends only on
them and $P_{\underbrace{y\ldots\ y}_{N_0+1}}=\partial_y^{N_0+1} M_P$.
Consequently calculating all the higher derivatives
$\partial_y^m M_P$ of the orders $m > N_0+1$,
we conclude that they depend only on the quantities
$P_{\underbrace{y\ldots\ y}_m}$, $m \leq N_0+1$.

The possibility to obtain
 $P_{\underbrace{x\ldots\ x}_{n}\underbrace{y\ldots\ y}_{N_0+1-n}} =
 \partial_x^n\partial^{N_0+1-n}_y M_P$,
$n \leq N_0+1$, will be proved by an auxiliary induction    over
$n$. The case  $n=0$ has been just considered.

Provided we have proved for all $n \leq n_0$ that
\begin{itemize}
\item[\rm a)] the principal derivatives $R^{(0)}_x$, $R^{(1)}_{xx}$, \ldots\ ,
$R^{(n)}_{\underbrace{x \ldots\ x}_{n+1}}$,
are already chosen in such a way that
 $P_{\underbrace{x\ldots\ x}_{k}\underbrace{y\ldots\ y}_{N_0+1-k}} =
 \partial_x^k\partial^{N_0+1-k}_y M_P$,
$k \leq n$, hold,
\item[\rm b)] in
\be
\begin{array}{l}
\displaystyle
 \partial_x^k\partial^{N_0+1-k}_y M_P =
    (-1)^{N_0} \Bigg(\partial_x^k \partial^{N_0+1-k}_y M_{K_0}\\
\displaystyle
     - \sum_{s=0}^{N_0+1} (-1)^s
   \frac{2 \partial_x^{k+1}R^{(s)}\partial_y^{N_0+2-k}R^{(s)}}
             {\big(R^{(s)}\big)^2} \\ +
\displaystyle
 F(R^{(q)}, \partial_x^mR^{(q)},  \partial_y^kR^{(q)},
     M_{K_0+q},\partial_x^r\partial_y^mM_{K_0+q}) \Bigg) ,
\end{array}
\ee{n0xy}
for $k \leq n$ in the terms collected in $F$ only the principal derivatives
 defined during the previous steps as well as
$\partial_x^r\partial_y^mM_{K_0+q}$ with $r \leq n$, $m+r \leq N_0$,
$K_0 +q < P$, are present,
\item[\rm c)] all the higher derivatives
$ \partial_x^m\partial^{k}_y M_P$, $m+k > N_0+1$, $m \leq n$,
and the already defined principal derivatives
$\partial_x^{s+1}R^{(s)}$, $s \leq n$,
depend (according to our construct of $M_P$) only on the choice of
 $P_{\underbrace{x\ldots\ x}_{p}\underbrace{y\ldots\ y}_{q}} $,
$p+q \leq N_0+1$, $p \leq m$.
\end{itemize}
we will make the inductive step
 $n=n_0+1$. Then in the rest $F$ only the already defined
principal derivatives of $R^{(s)}$  and
$\partial_x^r\partial_y^mM_{K_0+q}$ with $r \leq n_0+1$, $m+r \leq N_0$, will
appear alongside with the still unknown
 $\partial_x^{n_0+1}\partial_y^mM_{K_0+q}$.

In order to determine these quantities we will
 make another imbedded induction
over $m$ using the \e s
 $P_{\underbrace{x\ldots\ x}_{n_0+1}\underbrace{y\ldots\ y}_{N-n_0-1}} =
 \partial_x^{n_0+1}\partial^{N-n_0-1}_y M_P$,
$N \leq N_0$. Indeed for
$m=0$ in
\be
 \partial_x^{n_0+1} M_P =
    (-1)^{N_0} \left(\partial_x^{n_0+1} M_{K_0}
     - \sum_{s=0}^{N_0+1} (-1)^s
   \frac{2 \partial_x^{n_0+2}R^{(s)}R^{(s)}_y}
             {\big(R^{(s)}\big)^2}  \right) + F
\ee{f8}
($n_0+1\leq N_0+1$) only the known quantities appear except
$\partial_x^{n_0+1}M_{K_0}$ which 
we may choose according to the main inductive
proposition in such a way that
 $P_{\underbrace{x\ldots\ x}_{n_0+1}} =
 \partial_x^{n_0+1} M_P|_{(0,0)}$ hold.
Differentiating (\ref{mks}) we find
$ \partial_x^{n_0+1} M_{K_0+q}|_{(0,0)}$, $K_0+q < P$
inductively over $q$.  The step of the
 imbedded   induction we perform as earlier
applying the operator $\partial_y^m$ to
 (\ref{f8}). Again according to the main inductive proposition
(valid for $M_{K_0}$)
$\partial_x^{n_0+1}\partial_y^{N_0-n_0} M_{K_0}$ depends only upon
$\partial_x^n\partial_y^m M_{K_0}$, $m+n \leq N_0$, $n \leq n_0+1$,
which due to (\ref{n0xy}) implies the same dependence of
the principal derivative
$\partial_x^{n_0+2}R^{n_0+1}$. The validity of a), b), c) is easily checked
now for $n= n_0+1$.

The auxiliary induction over $n$ proves that we can
consecutively choose definite values of $R^{(0)}_{x}$, 
$R^{(1)}_{xx}$,  \ldots , $\partial_x^{N_0+2}R^{(N_0+1)}$ in
such a way that (\ref{sp}) hold for $N= N_0+1$ and (\ref{sp})
holds also for $N \leq N_0$ after an appropriate choice of 
$\partial_x^m\partial_y^m M_{K_0}$, $m+r \leq N_0$, which is
possible due to the inductive proposition. The higher derivatives
$\partial_x^m\partial_y^m M_{K_0}$, $m+r > N_0+1$, depend only
upon the lower order ones as indicated in the inductive proposition.
Therefore the main inductive proposition as well as Theorem are proved.

\end{document}